\documentclass[preprint]{aastex}
\singlespace

\usepackage{psfig}
 
\begin{document} 
\title{The ATNF Pulsar Catalogue}
\author{R. N. Manchester, G. B. Hobbs, A. Teoh and M. Hobbs}
\affil{Australia Telescope National Facility, CSIRO, PO Box 76,
        Epping NSW 1710, Australia}
\email{Dick.Manchester@csiro.au}

\begin{abstract} 
We have compiled a new and complete catalog of the main properties of
the 1509 pulsars for which published information currently exists. The catalog
includes all spin-powered pulsars as well as anomalous X-ray pulsars
and soft gamma-ray repeaters showing coherent pulsed emission, but
excludes accretion-powered systems. References are given for all data
listed. We have also developed a new web interface for accessing and
displaying either tabular or plotted data with the option of selecting
pulsars to be displayed via logical conditions on parameter
expressions. The web interface has an ``expert'' mode giving access to a
wider range of parameters and allowing the use of custom
databases. For users with locally installed software and database on unix or
linux systems, the catalog may be accessed from a command-line
interface. C-language functions to access specified parameters are
also available. The catalog is updated from time to time to include new
information. 
\end{abstract}

\keywords{pulsars: general --- astronomical databases: catalogs}

\section{Introduction}
Since the discovery of the first pulsar, announced by \citet{hbp+68},
the number of known pulsars has grown to more than 1500. About half of
these have been discovered in the past few years by surveys carried
out using the multibeam receiver on the Parkes 64-m radio telescope
\citep{mlc+01,ebvb01,mhl+02,kbm+03,hfs+04}. Although most known
pulsars were discovered at radio frequencies, recent X-ray
observations have resulted in the discovery of a number of pulsars
\citep[e.g.,][]{hh92,mgz+98}; some of these have been subsequently
detected at radio wavelengths, others have not. All of these pulsars
are powered by the rotational kinetic energy of the underlying neutron
star.

There exists another group of pulsars, detected at X-ray and gamma-ray
wavelengths, which are evidently isolated neutron stars, spinning down
in much the same way as ordinary pulsars, but where the pulsed
emission is too luminous to be powered by the spin-down energy
\citep[e.g.,][]{kds+98,tkk+98} These pulsars, known as anomalous X-ray
pulsars (AXPs) or soft gamma-ray repeaters (SGRs), have long pulse
periods but very rapid spin-down rates implying ultra-strong magnetic
fields. The X-ray emission in these so-called ``magnetars'' is
believed to be powered by relaxation of the strong magnetic fields
\citep[e.g.,][]{td96a}. Because these systems are similar in most
respects to ordinary pulsars, we have included them in the catalog. In
contrast, accretion-powered X-ray pulsars are quite different, with
pulse periods covering a wide range from milliseconds to minutes and
often being quite unstable on short timescales. Several hundred of
these systems are known and catalogs of their properties exist
\citep[e.g.,][]{bcc+97}, so we decided not to include them in the
present catalog.

The last published pulsar catalog \citep{tml93} contained 558 radio
pulsars. Various groups have privately maintained and updated this
catalog over the past decade.  However, with the recent rapid increase
in the number of known pulsars, even the best of these was seriously
incomplete.  We have taken the version of the catalog maintained by
the ATNF, Jodrell Bank Observatory and by other members of our pulsar
collaboration as the basis for a new catalog. We have extensively
searched the pulsar literature over the past decade for details of new
and previously known pulsars and built up a database containing full
bibliographic information. The catalog currently contains data for
1509 pulsars.

To make the catalog available to the wider community, we have
developed a versatile web interface which allows both tabulation and
plotting of selected data. A total of 67 different pulsar parameters
are pre-defined. Custom variables may be defined as functions of
pulsar parameters and tabulated or plotted. Data can be selected using
logical conditions on parameter expressions or distance from a
specified location. The web interface also has an ``expert'' mode of
operation which allows access to a wider range of parameters and the
ability to use one or more custom databases, either replacing or
merged with the public database. All functions of the web interface
except plotting are available using a command-line interface to the
catalog program {\sc psrcat}. This interface has been tested on
Macintosh OS and various flavours of linux and unix
systems. C-language functions which extract parameters from the
database are also available.

\section{The catalog database}
The catalog database is an ascii text file with a keyword--value
structure based on the system originally developed at the University
of Massachusetts \citep{mt72}. In addition to the keyword and value,
most observed parameters have additional fields for the error and
reference key. The available parameters, their keywords and units are listed in
Table~\ref{tb:basic} for basic parameters and
Table~\ref{tb:expert} for expert-mode parameters. Table~\ref{tb:db}
shows the format of the database entry for a representative (binary)
pulsar. Errors refer to the last quoted digit of the associated
parameter. Data for a given pulsar must start with the pulsar name (PSRB if it
exists, otherwise PSRJ) and must be terminated with a line beginning with
``@'', but otherwise the parameter order is immaterial. 

\begin{deluxetable}{ll}
\tabletypesize{\footnotesize}
\tablecaption{Basic Parameters\label{tb:basic}}
\tablehead{
\colhead{Keyword} & \colhead{Parameter Description}}
\startdata
\multicolumn{2}{l}{\bf Name and Position Parameters:} \\
Name  &  Pulsar name. The B1950 name if it exists, otherwise the J2000 name. \\
JName  &  Pulsar name based on J2000 coordinates\\
RAJ  &  Right ascension (J2000) (hh:mm:ss.s)\\
DecJ  &  Declination (J2000) (+dd:mm:ss)\\
PMRA  &  Proper motion in the right ascension direction (mas yr$^{-1}$)\\
PMDec  &  Proper motion in declination (mas yr$^{-1}$)\\
PX  &  Annual parallax (mas)\\
PosEpoch  &  Epoch at which the position is measured (MJD)\\
ELong  &  Ecliptic longitude (deg.)\\
ELat  &  Ecliptic latitude (deg.)\\
PMElong  &  Proper motion in the ecliptic longitude direction (mas yr$^{-1}$)\\
PMElat  &  Proper motion in ecliptic latitude (mas yr$^{-1}$)\\
GL  &  Galactic longitude (deg.)\\
GB  &  Galactic latitude (deg.)\\
RAJD  &  Right ascension (J2000) (deg.)\\
DecJD  &  Declination (J2000) (deg.)\\[10pt]
\multicolumn{2}{l}{\bf Timing and Profile Parameters:}\\
P0  &  Barycentric period of the pulsar (s)\\
P1  &  First time derivative of barycentric period \\
F0  &  Barycentric rotation frequency (Hz)\\
F1  &  First time derivative of barycentric rotation frequency (s$^{-2}$) \\
F2  &  Second time derivative of barycentric rotation frequency (s$^{-3}$) \\
F3  &  Third time derivative of barycentric rotation frequency (s$^{-4}$) \\
PEpoch  &  Epoch of period or frequency (MJD)\\
DM  &  Dispersion measure (cm$^{-3}$ pc)\\
DM1  &  First time derivative of dispersion measure (cm$^{-3}$ pc yr$^{-1}$) \\
RM  &  Rotation measure (rad m$^{-2}$) \\
W50  &  Width of pulse at 50\% of peak (ms).\tablenotemark{a}\\
W10  &  Width of pulse at 10\% of peak (ms).\tablenotemark{a}\\
Tau\_sc  &  Temporal broadening of pulses at 1 GHz due to interstellar scattering (s)\\
S400  &  Mean flux density at 400 MHz (mJy)\\
S1400  &  Mean flux density at 1400 MHz (mJy)\\
SPINDX  &  Measured spectral index\\[10pt]
\multicolumn{2}{l}{\bf Binary System Parameters:}\\
Binary  &  Binary model\tablenotemark{b}\\ 
T0  &  Epoch of periastron (MJD)\\
PB  &  Binary period of pulsar (days)\\
A1  &  Projected semi-major axis of pulsar orbit, $a_1\sin i$ (s)\\
OM  &  Longitude of periastron, $\omega$ (deg.)\\
Ecc  &  Eccentricity, $e$\\
Tasc  &  Epoch of ascending node (MJD)\\
Eps1  &  $e\sin\omega$ - ELL1 binary model\\
Eps2  &  $e\cos\omega$ - ELL1 binary model\\
MinMass  &  Minimum companion mass ($i=90\degr$, M$_{\rm NS} = 1.35$ M$_{\sun}$)\\
MedMass  &  Median companion mass ($i=60\degr$)\\[10pt]
\multicolumn{2}{l}{\bf Distance Parameters:}\\
Dist  &  Best estimate of the pulsar distance (kpc)\\
Dist\_DM  &  Distance based on the Taylor \& Cordes (1993)
 electron density model.\tablenotemark{c} \\
DMsinb  &  `Vertical' component of DM: ${\rm DM} \sin{\rm GB}$ (cm$^{-3}$ pc)\\
ZZ  &  Distance from the Galactic plane, based on Dist\\
XX  &  X-Distance in X-Y-Z Galactic coordinate system (kpc)\\
YY  &  Y-Distance in X-Y-Z Galactic coordinate system (kpc)\\[10pt]
\multicolumn{2}{l}{\bf Associations and Survey Parameters:}\\
Assoc  &  Names of associated objects\tablenotemark{d}\\
Survey  &  Surveys that detected the pulsar (discovery survey first).\tablenotemark{e}\\ 
OSurvey  &  Surveys that detected the pulsar as binary-encoded integer.\tablenotemark{e}\\ 
Date  &  Date of discovery publication.\\
Type  &  Type codes for the pulsar.\tablenotemark{f} \\
NGlt  &  Number of glitches observed for the pulsar\\[10pt]
\multicolumn{2}{l}{\bf Derived Parameters:}\\
R\_Lum  &  Radio luminosity at 400 MHz (mJy kpc$^{2}$) \\
R\_Lum14  &  Radio luminosity at 1400 MHz (mJy kpc$^{2}$) \\
Age  &  Characteristic age  (yr) \\
BSurf  &  Surface dipole magnetic flux density (G)\\
Edot  &  Spin down energy loss rate (erg s$^{-1}$)\\
Edotd2  &  Energy flux at the Sun (erg s$^{-1}$ kpc$^{-2}$)\\
PMTot  &  Total proper motion (mas yr$^{-1}$)\\
VTrans  &  Transverse velocity - based on Dist (km s$^{-1}$)\\
P1\_i  &  Period derivative corrected for Shklovskii effect\\
Age\_i  &  Characteristic age from P1\_i (yr)\\
BSurf\_i  &  Surface magnetic dipole from P1\_i (G)\\
Edot\_i  &  Spin down energy loss rate from P1\_i (erg s$^{-1}$)\\
B\_LC  &  Magnetic field at light cylinder (G)\\
\enddata
\tablenotetext{a}{Pulse widths are a function of both observing frequency 
and observational time resolution, so quoted widths are indicative
only.}
\tablenotetext{b}{Normally a binary model defined by the pulsar
timing program
\anchor{http://www.atnf.csiro.au/research/pulsar/timing/tempo}{TEMPO}} 
\tablenotetext{c}{In ``Long'' or ``Publication quality'' modes,
 lower limits from the distance model are preceded by a `+' sign.}
\tablenotetext{d}{See Table~\ref{tb:assoc}}
\tablenotetext{e}{See Table~\ref{tb:surveys}}
\tablenotetext{f}{See Table~\ref{tb:types}}
\end{deluxetable}

\begin{deluxetable}{ll}
\tabletypesize{\footnotesize}
\tablecaption{Expert Parameters\label{tb:expert}}
\tablehead{
\colhead{Keyword} & \colhead{Parameter Description}}
\startdata
\multicolumn{2}{l}{\bf Name and Position Parameters:} \\
Bname  &  Pulsar name based on B1950 coordinates \\
Alias  &  Alternative name \\
PML  &  Proper motion in the Galactic longitude direction (mas yr$^{-1}$) \\
PMB  &  Proper motion in Galactic latitude (mas yr$^{-1}$) \\[10pt]
\multicolumn{2}{l}{\bf Timing and Profile Parameters:} \\
F4  &  Fourth time derivative of barycentric rotation frequency (s$^{-5}$) \\
F5  &  Fifth time derivative of barycentric rotation frequency (s$^{-6}$) \\ 
F6  &  Sixth time derivative of barycentric rotation frequency (s$^{-7}$) \\ 
F7  &  Seventh time derivative of barycentric rotation frequency (s$^{-8}$) \\ 
F8  &  Eighth time derivative of barycentric rotation frequency (s$^{-9}$) \\
F9  &  Ninth time derivative of barycentric rotation frequency (s$^{-10}$) \\ 
FA  &  Tenth time derivative of barycentric rotation frequency (s$^{-11}$) \\ 
FB  &  Eleventh time derivative of barycentric rotation frequency (s$^{-12}$) \\
FC  &  Twelfth time derivative of barycentric rotation frequency (s$^{-13}$) \\ 
DM2  &  Second time derivative of dispersion measure (cm$^{-3}$ pc yr$^{-2}$) \\ 
DM3  &  Third time derivative of dispersion measure (cm$^{-3}$ pc yr$^{-3}$) \\ 
DM4  &  Fourth time derivative of dispersion measure (cm$^{-3}$ pc yr$^{-4}$) \\
DM5  &  Fifth time derivative of dispersion measure (cm$^{-3}$ pc yr$^{-5}$) \\ 
DM6  &  Sixth time derivative of dispersion measure (cm$^{-3}$ pc yr$^{-6}$) \\ 
DM7  &  Seventh time derivative of dispersion measure (cm$^{-3}$ pc yr$^{-7}$) \\ 
DM8  &  Eighth time derivative of dispersion measure (cm$^{-3}$ pc yr$^{-8}$) \\ 
DM9  &  Ninth time derivative of dispersion measure (cm$^{-3}$ pc yr$^{-9}$) \\
Interim  &  Interim timing solution \\
S600  &  Mean Flux Density at 600 MHz (mJy)  \\ 
S925  &  Mean Flux Density at 925 MHz (mJy)  \\ 
S1600  &  Mean Flux Density at 1600 MHz (mJy)  \\
SI414  &  Spectral index between 400 and 1400 MHz  \\[10pt]
\multicolumn{2}{l}{\bf Binary Parameters:} \\
OMDOT  &  Periastron advance (deg yr$^{-1}$)  \\
PBDOT  &  First time derivative of binary period   \\
A1DOT  &  Rate of change of projected semi-major axis  \\
ECCDOT  &  Rate of change of eccentricity (s$^{-1}$) \\
GAMMA  &  Relativistic time dilation term (s)  \\
T0\_2  &  Epoch of periastron [2nd orbit] (MJD)  \\
PB\_2  &  Binary period of pulsar [2nd orbit] (days)  \\
A1\_2  &  Projected semi-major axis of orbit [2nd orbit] (s)  \\
OM\_2  &  Longitude of periastron [2nd orbit] (deg)  \\
OMDOT\_2  &  Periastron advance [2nd orbit] (deg yr$^{-1}$)  \\
ECC\_2  &  Eccentricity [2nd orbit]  \\
PBDOT\_2  &  1st time derivative of binary period [2nd orbit]  \\
T0\_3  &  Epoch of periastron [3rd orbit] (MJD)  \\
PB\_3  &  Binary period of pulsar [3rd orbit] (days)  \\
A1\_3  &  Projected semi-major axis of orbit [3rd orbit] (s)  \\
OM\_3  &  Longitude of periastron [3rd orbit] (deg)  \\
OMDOT\_3  &  Periastron advance [3rd orbit] (deg yr$^{-1}$)  \\
ECC\_3  &  Eccentricity [3rd orbit]  \\
PBDOT\_3  &  1st time derivative of binary period [3rd orbit]  \\
PPNGAMMA  &  PPN parameter gamma  \\
SINI  &  Sine of inclination angle $i$  \\
SINI\_2  &  Sine of inclination angle [2nd orbit] \\
SINI\_3  &  Sine of inclination angle [3rd orbit] \\
MTOT  &  Total system mass (M$_{\sun}$) \\
M2  &  Companion mass (M$_{\sun}$) \\
M2\_2  &  Companion mass [2nd orbit] (M$_{\sun}$) \\
M2\_3  &  Companion mass [3rd orbit] (M$_{\sun}$) \\
DTHETA  &  Relativistic deformation of the orbit  \\
XOMDOT  &  Rate of periastron advance minus GR prediction (deg yr$^{-1}$)  \\
XPBDOT  &  Rate of change of orbital period minus GR prediction   \\
DR  &  Relativistic deformation of the orbit  \\
A0  &  Aberration parameter A0  \\
B0  &  Aberration parameter B0 (s)  \\
BP  &  Tensor multi-scalar parameter $\beta^{\prime}$  \\
BPP  &  Tensor multi-scalar parameter $\beta^{\prime\prime}$  \\
MASSFN  &  The pulsar mass function (M$_{\sun}$) \\
UPRMASS  &  90\% confidence upper companion mass limit, $i=26\degr$ (M$_{\sun}$) \\
MINOMDOT  &  Minimum OMDOT, assuming $i=90\degr$ and M$_{\rm NS}$ = 1.4M$_{\sun}$ (deg yr$^{-1}$) \\[10pt]
\multicolumn{2}{l}{\bf Other Timing Parameters:} \\
TRES  &  RMS timing residual ($\mu s$)\tablenotemark{a}  \\
NTOA  &  Number of TOAs in timing fit\tablenotemark{a}  \\
START  &  Epoch of start of fit (MJD)\tablenotemark{a}  \\
FINISH  &  Epoch of end of fit (MJD)\tablenotemark{a}  \\
CLK  &  Terrestrial time standard\tablenotemark{a}  \\
EPHEM  &  Solar system ephemeris\tablenotemark{a}  \\
TZRMJD  &  Reference TOA\tablenotemark{a}~(MJD)  \\
TZRFRQ  &  Frequency of reference TOA\tablenotemark{a}~(MHz)  \\
TZRSITE  &  One-letter observatory code for reference TOA\tablenotemark{a}   \\
NSPAN  &  Polyco span\tablenotemark{a}~(min)  \\
NCOEF  &  Number of coefficients in polyco\tablenotemark{a}  \\
GLEP  &  Epoch of glitch  \\
GLPH  &  Phase increment at glitch  \\
GLF0  &  Permanent pulse frequency increment at glitch  \\
GLFI  &  Permanent frequency derivative increment at glitch \\
GLF0D  &  Decaying frequency increment at glitch   \\
GLTD  &  Time constant for decaying frequency increment  \\[10pt]
\multicolumn{2}{l}{\bf Distance Parameters:} \\
Dist\_DM1  &  Distance based on NE2001 model (kpc) \\
Dist1  &  Best estimate of pulsar distance using Dist\_DM1 as default \\
Dist\_AMN  &  Lower limit on distance based on association or HI absorption (kpc) \\
Dist\_AMX  &  Upper limit on distance based on association or HI absorption (kpc) \\
Dist\_A  &  Distance based on association or HI absorption (kpc) \\[10pt]
\multicolumn{2}{l}{\bf User-defined Parameters:} \\
PAR1  &  A user-defined catalog entry \\
PAR2  &  A user-defined catalog entry \\
PAR3  &  A user-defined catalog entry \\
PAR4  &  A user-defined catalog entry \\[10pt]
\enddata
\tablenotetext{a}{Available in command-line version only.} 
\end{deluxetable}

\begin{deluxetable}{llll}
\tabletypesize{\small}
\tablecaption{A representative database entry\label{tb:db}}
\tablehead{
\colhead{Keyword} & \colhead{Value}& \colhead{Error}& \colhead{Ref. Key}}
\startdata
PSRJ    &  J1435$-$6100           &      &  clm+01 \\      
RAJ     &  14:35:20.2765          &   5  &  clm+01 \\       
DECJ    &  $-$61:00:57.956        &   7  &  clm+01 \\       
F0      &  106.97507197376        &   8  &  clm+01 \\       
F1      &  $-$2.80E$-$16          &   5  &  clm+01 \\       
PEPOCH  &  51270.000              &      &         \\       
DM      &  113.7                  &   6  &  clm+01 \\       
BINARY  &  ELL1                   &      &         \\       
TASC    &  51270.6084449          &   6  &  clm+01 \\       
PB      &  1.3548852170           &   18 &  clm+01 \\       
A1      &  6.184023               &   4  &  clm+01 \\       
EPS1    &  1.9E$-$6               &   12 &  clm+01 \\       
EPS2    &  1.03E$-$5              &   15 &  clm+01 \\       
START   &  50939.602              &      &         \\       
FINISH  &  51856.205              &      &         \\       
TRES    &  83.97                  &      &         \\       
NTOA    &  93                     &      &         \\       
CLK     &  UNCORR                 &      &         \\       
EPHEM   &  DE200                  &      &         \\       
TZRMJD  &  51293.55635374447232   &      &         \\       
TZRFRQ  &  1374.000               &      &         \\       
TZRSITE &  7                      &      &         \\       
S1400   &  0.25                   &   4  &  mlc+01 \\       
W50     &  1.1                    &      &  mlc+01 \\       
DIST\_DM &  3.25                  &      &  tc93   \\       
DIST\_DM1&  2.16                  &      &  cl02   \\       
SURVEY  &  pksmb                  &      & 	   \\ 
\verb#@------------#
\enddata
\end{deluxetable}

All data values have an associated reference key for the source of the
value and its error. The keys refer to a {\sc Bibtex} bibliography
database and are used to create a bibliography which currently has
more than 360 entries. The complete bibliography may be listed from
both the command-line and web interfaces.

Up to about 1993, pulsars were given a name based on their coordinates
in the Besselian 1950 system. At that time the J2000 coordinate system
was introduced and, following this, most pulsars were given names
based on their J2000 coordinates. For consistency, pulsars with B1950
names have been given a new name based on their J2000
coordinates. However, recently discovered pulsars are not given a
B1950 name. In accordance with
\anchor{http://cdsweb.u-strasbg.fr/iau-spec.html}{IAU
specifications}\footnote{See
http://cdsweb.u-strasbg.fr/iau-spec.html}, names explicitly include
the equinox letter, e.g., PSR~B0833$-$45 or PSR~J0835$-$4510. Note
however, that positions can only be given in J2000 or ecliptic
coordinates; B1950 coordinates are not supported. The parameter {\sf
PosEpoch} is the epoch of the position, expressed as a Modified Julian
Day (MJD = JD $-$ 2400000.5). If this parameter is not explicitly in
the database, it is taken to be the epoch of the pulse period ({\sf
PEpoch}).
 
Pulse timing parameters are closely related to the timing analysis
program \anchor{http://www.atnf.csiro.au/research/pulsar/tempo}{{\sc
tempo}}\footnote{See http://www.atnf.csiro.au/research/pulsar/tempo}.
Binary parameters, in particular, depend on the exact definition in
this program. The \citet{bt76} BT binary model is the most commonly
used description. However for binary systems with circular or
near-circular orbits the ELL1 model \citep{wex00} is more appropriate
and, for binary systems where relativistic effects are important, the
DD model \citep{dd86} provides a more exact treatment. Other binary
models are also supported -- see the {\sc TEMPO} documentation for
more details.

Some pulsars, especially young pulsars, occasionally suffer a sudden
decrease in pulse period, commonly known as a ``glitch''. The
parameter {\sf NGlt} is the total number of observed glitches in a
given pulsar. There is provision in expert mode for entering
and accessing parameters for one glitch, based on the glitch model in {\sc
TEMPO}. These parameters are defined by:
\begin{equation}
\label{eq:glitch}
\nu(t) = \nu_0(t) + \Delta\nu_p + \Delta\dot\nu_p t + 
      \Delta\nu_d \exp(-t/\tau_d),
\end{equation}
where $\nu$ is the pulse frequency, $\nu_0$ is its value at the glitch
epoch ({\sf GLEP}, $t=0$) extrapolated from pre-glitch data, $\Delta\nu_p$ ({\sf
GLF0}) and $\Delta\dot\nu_p$ ({\sf GLF1}) are the permanent changes in
$\nu$ and $\dot\nu$ at the time of the glitch, $\Delta\nu_d$ ({\sf
GLF0D}) is the decaying part of the frequency increment at the time of
the glitch, and $\tau_d$ ({\sf GLTD}) is the decay timescale. For
$t<0$, $\Delta\nu_p$, $\Delta\dot\nu_p$ and $\Delta\nu_d$ are all
zero. {\sc TEMPO} also provides a pulse phase increment at $t=0$ ({\sf
GLPH}) to allow for error in the assigned glitch epoch. 

A table of the basic glitch parameters for each pulsar known to glitch
({\sf NGlt} $> 0$) may be accessed by clicking on the pulsar
name. Parameters listed are the glitch epoch, the fractional change in
pulse frequency, $(\Delta\nu_p + \Delta\nu_d)/\nu_0$, and the
fractional change in frequency derivative $(\Delta\dot\nu_p -
\Delta\nu_d/\tau_d)/\dot\nu_0$, where $\dot\nu_0$ is the value of
$\dot\nu$ at $t=0$, extrapolated from the pre-glitch data, and their
estimated errors. Note that the simple exponential decay given by
Equation~\ref{eq:glitch} does not fully describe the post-glitch
behaviour in many cases. Note also that, if the measured value of
$\Delta\dot\nu$ is simply based on the observed pre- and post-glitch values of
$\dot\nu$ or if the single exponential decay model is not accurate,
the derived value may under-estimate the actual change in $\dot\nu$ at
the time of the glitch.

The pulsar distance $d$ ({\sf Dist}) depends on other catalog
parameters and is not itself a catalog entry. The default value is
that derived from the dispersion measure ({\sf DM}) using the
\citet{tc93} model for the Galactic distribution of free electrons,
i.e., {\sf Dist} = {\sf Dist\_DM}. However, if there is a measured
annual parallax ({\sf PX}), this takes precedence: $d=1/\pi$ where
$\pi$ is the parallax. Next in priority order is a distance estimate
({\sf Dist\_A}) based on an association with another object (e.g.,
globular cluster or supernova remnant) or measurements of absorption
by neutral hydrogen combined with a model for differential rotation of
the Galaxy. The classes of associated objects given in the catalog
(with keyword {\sf Assoc}) are listed in Table~\ref{tb:assoc}. If {\sf
Dist\_A} exists, {\sf Dist} is set equal to that. If there are only
distance limits ({\sf Dist\_AMN}, {\sf Dist\_AMX}), then {\sf Dist} is
set equal to the DM-derived distance if it lies between these limits
or to the nearest limit if it doesn't. {\sf Dist\_A} and the limits
{\sf Dist\_AMN} and {\sf Dist\_AMX} are available in expert mode.
{\sf Dist\_DM1}, a distance estimate based on the NE2001 Galactic
electron-density model \citep{cl02} and the associated {\sf Dist1} are
also available in expert mode. The Galactocentric coordinate system
({\sf XX, YY, ZZ}) is right-handed with the Sun at (0.0, 8.5 kpc, 0.0)
and the {\sf ZZ} axis directed toward the north Galactic pole.

\begin{deluxetable}{ll}
\tabletypesize{\small}
\tablecaption{Association Types\label{tb:assoc}}
\tablehead{
\colhead{Label} & \colhead{Description}}
\startdata
EXGAL & External galaxy \\
GC & Globular cluster \\
GRS & Gamma-ray source \\
OPT & Optical identification \\
SNR & Supernova remnant \\
XRS & X-ray source \\
\enddata
\end{deluxetable}

The major pulsar surveys and their associated labels are listed in
Table~\ref{tb:surveys}. The keyword {\sf Survey} gives labels for
those surveys which have detected a pulsar, with the discovery survey
listed first. All but 150 of the nearly 1500 pulsars have been
discovered in one of the major surveys listed; the remainder are
listed under ``misc''. The parameter {\sf OSurvey} is an octal-coded
integer with each survey associated with a particular bit of the
binary word.

\begin{deluxetable}{llrrr}
\tabletypesize{\small}
\tablecaption{Pulsar Surveys\label{tb:surveys}}
\tablehead{
\colhead{Survey} & \colhead{Survey} & \colhead{Octal} &
\colhead{Number} & \colhead{Number}\\
\colhead{Label} & \colhead{Name} & \colhead{Code} &
\colhead{Detected} & \colhead{Discovered}}
\startdata
ar1 & Arecibo Survey 1 & 4 & 49 & 41 \\
ar2 & Arecibo Survey 2 & 400 & 24 & 6 \\
ar3 & Arecibo Survey 3 & 2000 & 63 & 25 \\
ar4 & Arecibo Survey 4 & 20000 & 87 & 62 \\
gb1 & Green Bank Northern Survey & 20 & 50 & 31 \\
gb2 & Princeton-NRAO Survey & 40 & 83 & 34 \\
gb3 & Green Bank Short-Period Survey & 200 & 86 & 20 \\
gb4 & Green Bank Fast Pulsar Survey & 10000 & 8 & 5 \\
jb1 & Jodrell Bank A Survey & 2 & 51 & 45 \\
jb2 & Jodrell Bank B Survey & 100 & 62 & 42 \\
misc & \nodata & 400000 & 150 & 150 \\
mol1 & 1st Molonglo Survey & 1 & 35 & 35 \\
mol2 & 2nd Molonglo Survey & 10 & 224 & 155 \\
pks1 & Parkes 20-cm Survey & 1000 & 100 & 46 \\
pks70 & Parkes Southern-Sky Survey & 4000 & 298 & 101 \\
pksmb & Parkes Multibeam Survey & 40000 & 880 & 592 \\
pksgc & Parkes Globular Cluster Survey & 200000 & 10 & 10 \\
swmb & Swinburne Multibeam Survey & 100000 & 170 & 69 \\
\enddata
\end{deluxetable}

Pulsar types are listed in Table~\ref{tb:types}. Types AXP, HE and NR
are explicitly listed in the catalogue with keyword {\sf Type}. All
pulsars in a binary system with a measured orbital period are listed under type BINARY,
and all pulsars which are not type NR are listed under type RADIO. 

\begin{deluxetable}{ll}
\tabletypesize{\small}
\tablecaption{Pulsar Types\label{tb:types}}
\tablehead{
\colhead{Label} & \colhead{Description}}
\startdata
AXP & Anomalous X-ray pulsar or pulsating soft gamma-ray repeater \\
BINARY & Pulsar with one or more stellar or planetary companions \\
HE & Spin-powered pulsar with pulsed emission from radio to infrared
or higher frequencies \\
NR & Spin-powered pulsar with pulsed emission only at infrared
or higher frequencies \\
RADIO & Pulsars with pulsed emission in the radio band \\
\enddata
\end{deluxetable}

\section{Derived Parameters}
Both the web and command-line versions of the program allow display of
various parameters derived from catalog parameters as listed
in Table~\ref{tb:basic}. The radio ``luminosities'' {\sf R\_Lum} and
{\sf R\_Lum14}, commonly used in pulsar evolution and
distribution studies, are simply defined as $Sd^2$, where $S$ is {\sf
S400} or {\sf S1400} (in mJy) for {\sf R\_Lum} and {\sf R\_Lum14},
respectively, and $d$ is the pulsar distance ({\sf Dist}) in kpc. The
pulsar characteristic age ({\sf Age}) is defined by 
\begin{equation}
\label{eq:age}
\tau_c = P/(2 \dot P),
\end{equation}
where $P$ is the pulsar period ({\sf P0}) and $\dot P$ is its first
time derivative ({\sf P1}). 

Based on pulsar spin-down due to magnetic-dipole radiation, the
surface dipole magnetic flux density {\sf BSurf} is conventionally
defined to be
\begin{equation}
\label{eq:bsurf}
B_s = \left(\frac{3Ic^3 P\dot P}{8\pi^2 R_N^6}\right)^{1/2} 
    = 3.2\times 10^{19} (P \dot P)^{1/2}\;{\rm G},
\end{equation}
where $I$ is the neutron-star moment of inertia, assumed to be $10^{45}$
g cm$^2$, $R_N$ is the neutron-star radius, taken to be $10^6$ cm, $c$
is the velocity of light and $P$ is the pulsar period in seconds
\citep{mt77}. For a pure dipole field with the magnetic axis
perpendicular to the rotation axis, this is the field strength at
the magnetic equator; the field strength at the magnetic pole is a
factor of two higher. The magnetic flux density at the light-cylinder
radius $R_{LC} = cP/(2\pi)$ ({\sf B\_LC}) is computed assuming a
dipole field:
\begin{equation}
\label{eq:blc}
B_{LC} = B_s (R_N/R_{LC})^3 = 3.0\times 10^8 P^{-5/2} \dot
P^{1/2}\;{\rm G}.
\end{equation}
The rate of loss of rotational kinetic energy ({\sf Edot}) is given by
\begin{equation}
\label{eq:edot}
\dot E = -I\Omega\dot\Omega = 4\pi^2 I \dot P P^{-3}\;{\rm erg\;s}^{-1}
\end{equation}
where $\Omega = 2\pi /P$. The parameter {\sf Edotd2} is $\dot E
d^{-2}$, where $d$ is the pulsar distance. This is proportional to
the spin-down energy flux at the Earth and is a good indicator of the
detectability of high-energy, particularly gamma-ray, pulsed
emission. 

Proper motions are expressed in milliarcseconds (mas) on the sky per
year and may be entered in either J2000 coordinates ($\mu_{\alpha},
\mu_{\delta}$) or ecliptic coordinates. The proper motion in the other
coordinate system is computed from the entered values. Proper motions
in Galactic coordinates are also available in expert mode. Galactic
proper motions are computed from the entered values and are corrected
for the effects of Galactic rotation assuming a flat rotation curve
with a rotation velocity of 225 km~s$^{-1}$ \citep[cf.][]{hla93}.  The
total proper motion ({\sf PMTot}) is given by
\begin{equation}
\label{eq:pmtot}
\mu = (\mu_{\alpha}^2 + \mu_{\delta}^2)^{1/2},
\end{equation}
also in mas yr$^{-1}$, and the corresponding transverse velocity ({\sf
VTrans}) is given by
\begin{equation}
\label{eq:vtrans}
v_T = \mu d.
\end{equation}
As first pointed out by \citet{shk70}, a large
transverse velocity can introduce a significant kinematic term into
observed period derivatives:
\begin{equation}
\label{eq:pdshk}
\dot P_s = v_T^2 P/(cd).
\end{equation}
The {\it intrinsic} period derivative ({\sf P1\_i}),
\begin{equation}
\label{eq:pdoti}
\dot P_i = \dot P - \dot P_s,
\end{equation}
is a better measure of the actual slow-down rate of the pulsar and can
be significantly less than the measured value, especially for nearby
millisecond pulsars. For example, for PSR J0437$-$4715, the measured
$\dot P$ is about $5.7\times 10^{-20}$ whereas $\dot P_i$ is just one
third of this value. Likewise, {\sf Age\_i}, {\sf BSurf\_i} and {\sf
Edot\_i}, derived with $\dot P$ replaced by $\dot P_i$, are better
measures of the actual values of these quantities. 

The catalog interfaces allow definition, listing and (for the
web interface) plotting of up to four ``custom'' parameters ({\sf C1}
 -- {\sf C4}), that is,
parameters which are algebraic combinations of other parameters
(including other custom parameters). These (and all other) entries are
case insensitive. Available operators and functions are listed in
Table~\ref{tb:alg}. 

\begin{deluxetable}{llllll}
\tabletypesize{\small}
\tablecaption{Valid algebraic operators and functions for parameter
expressions\label{tb:alg}}
\tablehead{
\multicolumn{2}{c}{Operators} & \multicolumn{2}{c}{Functions} &
\multicolumn{2}{c}{Functions}
}
\startdata     
+  & addition  & acos($a$) & inverse cosine & sin($a$)   & sine of
angle in radians \\
$-$ & subtraction & asin($a$) & inverse sine   & sind($a$)  & sine of
angle in degrees \\
$\ast$  & multiplication & atan($a$) & inverse tangent & sinh($a$) & hyperbolic
sine \\
/  & division  & atan2($a,b$) & inverse tangent & sqr($a$) & square \\
$\ast\ast$ & Raise to power & cos($a$)  & cosine of angle in radians & sqrt($a$) &
square root \\
=  & assignment & cosd($a$) & cosine of angle in degrees & tan($a$) &
tangent of angle in radians \\
 & & cosh($a$) & hyperbolic cosine & tand($a$) & tangent of angle in
degrees \\
 & & exp($a$) & exponential & tanh($a$) & hyperbolic tangent \\
 & & ln($a$) & logarithm to base 2 & fabs($a$) & absolute value \\
 & & log($a$) & logarithm to base 10 & fmod($a,b$) & modulus of $a$ with
respect to $b$ \\
 & & log10($a$) & logarithm to base 10 & & \\
\enddata
\end{deluxetable}
  
Updates to the public database are made from time to time to correct any
errors and to include recently published data. The database file is
maintained under Concurrent Versions System (CVS)\footnote{See
http://www.cvshome.org/} control; the CVS version number of the
current file is displayed on the web interface and may be accessed
from the command-line interface.

\section{The web interface}
The main user interface to the catalog is provided by the
interactive \anchor{http://www.atnf.csiro.au/research/pulsar/psrcat}{web
page}\footnote{http://www.atnf.csiro.au/research/pulsar/psrcat}. This web
page provides access to most catalog parameters and to a range of
derived parameters, with facilities for both tabular and plotted
outputs. An extensive tutorial on the operation of the web interface
may be accessed either from a link at the top of the main page (which
creates a new browser page) or via links to individual sections of the
tutorial.  Documentation on parameter definitions and units can be
accessed either by a link at the top of the main page or, for
individual parameters, by clicking on the parameter name.

Parameters for tabular output may be selected from the displayed
list. Output values are typically of variable length, but all consist
of a single ascii string or number with no spaces. By default, null
values are represented by an asterisk, but it is possible for the user
to select a null character or string. These properties facilitate
free-format reading of tabular values with a space delimiter.  The
list may be sorted in either ascending or descending order by any
parameter, ascii or numeric, with a default of the J2000 name. It is
possible to select a ``No header'' option which omits the column
headings and also the space after every fifth line. This facilitates
selecting and pasting of tabular output into a text editor for use in
other applications.

Five different output formats are available:
\begin{enumerate}
\item {\bf Short without errors:} Lists parameters with a fixed format and a precision which is
often less than the available precision but more than adequate for
most applications needing input data. No errors or reference keys are
listed.
\item {\bf Short with errors:} Identical to ``short'' except
that, when available, errors are listed in exponential notation.
\item {\bf Long with last-digit error:} Gives all values to
the full available precision, lists the error in the last
quoted digit and the reference key for each data value. 
\item {\bf Long with error:} Similar to ``Long with last-digit
  error'' except that the error is quoted in exponential notation.
\item {\bf Publication quality:} Similar to ``Long with last-digit error''
  except that the error is given in parentheses at the end of the
  value and the reference keys are collected on the right-hand side of
  the line.
\end{enumerate}
 
\begin{figure}
\centerline{\psfig{file=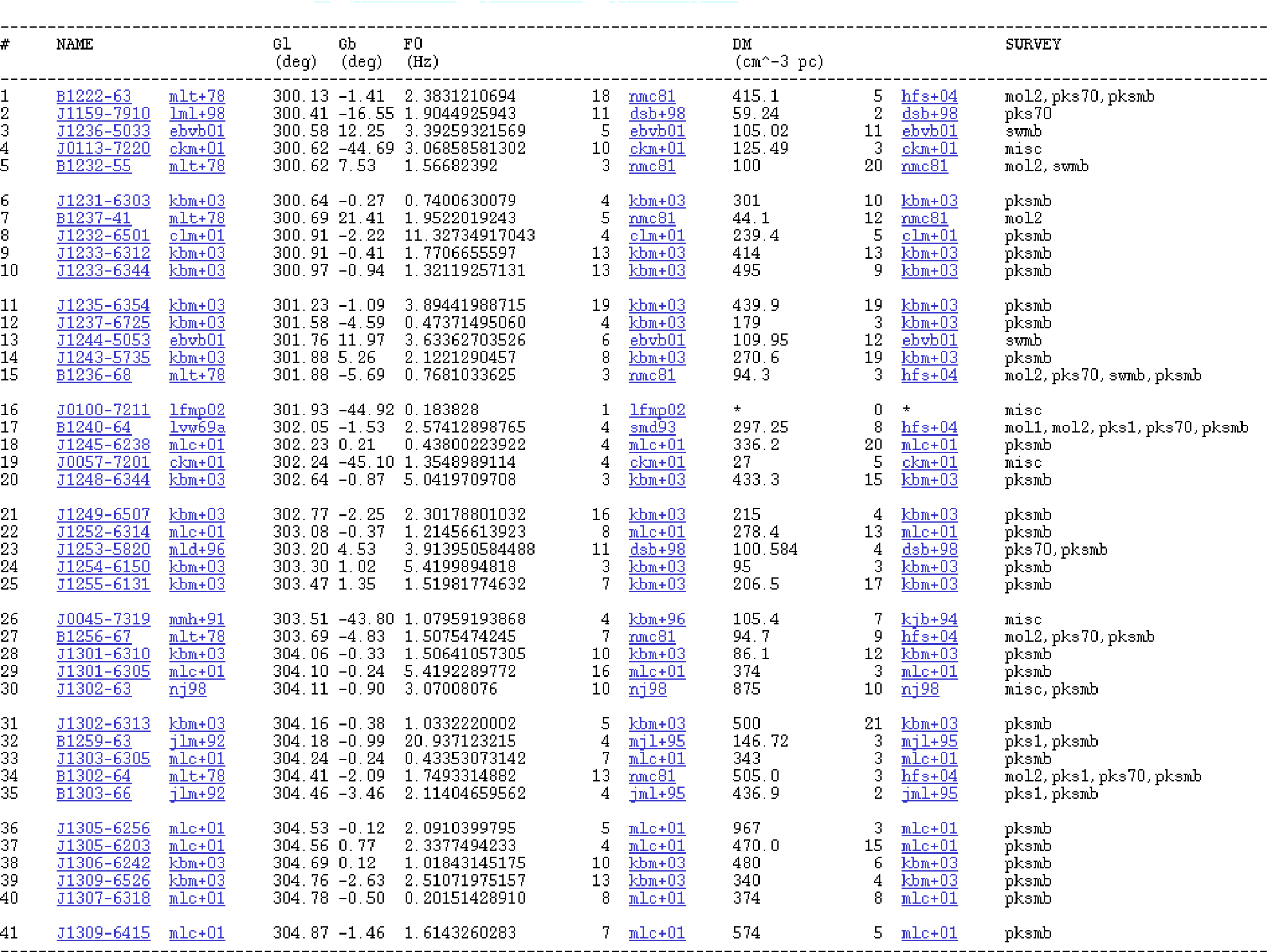,width=160mm}}
\caption{A typical tabular output from the {\sc psrcat} web interface
in the (default) long format with last-digit errors. This list was
limited to pulsars with Galactic longitude $l$ in the range $300\degr$ to
$305\degr$ and sorted in order of increasing $l$. Note the ``null''
character for the unmeasured dispersion measure for the AXP
J0100-7211.}\label{fg:tbl}
\end{figure}

Fig.~\ref{fg:tbl} shows a small segment of a typical tabular output in
the default ``Long with last-digit error'' format. Reference keys are
all linked to the appropriate part of the bibliography database giving
full reference information for the relevant publication. The reference
associated with a pulsar name is to the paper in which the discovery
of the pulsar was announced. The pulsar name itself is linked to the
European Pulsar Network
\anchor{http://www.mpifr-bonn.mpg.de/div/pulsar/data/archive.html}{web
page}\footnote{http://www.mpifr-bonn.mpg.de/div/pulsar/data/archive.html}
which gives spectra and mean-pulse polarization profiles for a large
number of pulsars, to the NASA Astrophysics Data System
\anchor{http://adsabs.harvard.edu/abstract_service.html}{(ADS)}\footnote{http://adsabs.harvard.edu/abstract\_service.html}
listing publications which refer to this pulsar, and to a table of
glitch parameters for this pulsar (if known to glitch).

\begin{figure}
\centerline{\psfig{file=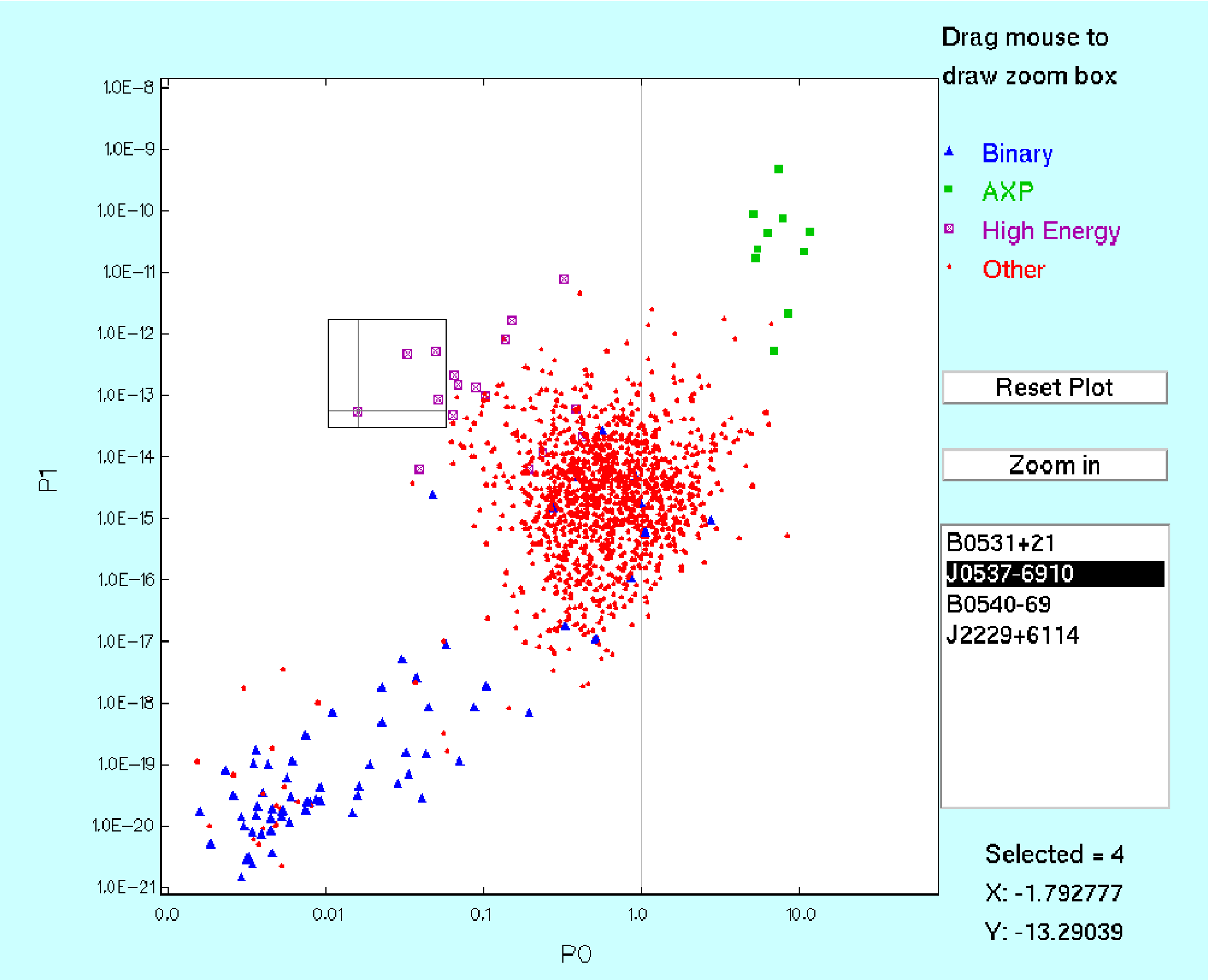,width=160mm}}
\caption{A plot of pulse period versus period derivative on
logarithmic scales produced by the {\sc psrcat} web
interface.}\label{fg:plot}
\end{figure}

The web interface also provides an interactive plotting facility. Any
(numeric) parameter may be plotted against any other parameter or as a
histogram on either linear or logarithmic scales. The main pulsar
types (binary, high-energy, AXP, other) are identified by different
symbols. Fig.~\ref{fg:plot} shows a typical plot. It is possible to
zoom into a selected region of the plot. Pulsars within a selected
region are identified by name in a side box and clicking on a name
draws crossed lines through the point for that pulsar. If only an
$x$-coordinate is entered, a histogram for the distribution of that
parameter is plotted. The number of boxes in the histogram can be
interactively adjusted and clicking on a box identifies the pulsars in
that box.

The list of pulsars for which data is tabulated or plotted may be
limited in various ways. Data can be displayed for just selected pulsars
by entering the pulsar names in a box. Wild-card entries with
``*'' and ``?'' are supported and both B1950 and J2000 names are
checked for a match. For example, ``b1933+1?'' will match PSRs
B1933+16, B1933+17 and B1933+15 whereas ``j004*+*'' will match PSRs
J0040+5716 and J0048+3412. Displayed data can also be limited by
logical conditions on parameter functions as well as several special
functions. Table~\ref{tb:logical} lists the available logical
operators and special functions. Finally, only pulsars within a nominated
(spherical) angle of a given position (expressed in celestial or
Galactic coordinates) can be listed or plotted. 

\begin{deluxetable}{llll}
\tabletypesize{\small}
\tablecaption{Logical operators and functions for pulsar selection\label{tb:logical}}
\tablehead{
\multicolumn{2}{c}{Logical Operators} & \multicolumn{2}{c}{Functions}
}
\startdata     
==  & equality  & exist($x$) & existence of value for parameter $x$ \\
!= & inequality & error($x$)  & returns value of error for parameter $x$ \\
! & logical not &  type($t$) & pulsar of type $t$  \\
\&\& & logical and &assoc($s$) & {\sf Assoc} contains string $s$ \\
$||$ & logical or & survey($s$) & {\sf Survey} contains string $s$ \\
$<$ & less than & discovery($s$) & discovery survey contains string $s$ \\
$<=$  & less than or equal to & ref($p,s$) & reference for parameter
$p$ contains string $s$ \\
$>$  & greater than  & hms($s$) & Convert from hr:min:sec string $s$ to decimal
degrees \\
$>=$ & greater than or equal to & dms($s$) & Convert from deg:min:sec
string $s$ to decimal degrees \\
\enddata
\end{deluxetable}

Parameters for one or more named pulsars can be output as a table
containing keywords, values (to full precision), and errors (in
exponential notation) in ``ephemeris'' format, that is, a line for
each parameter. Three output options are provided: short mode lists
those parameters which are normally needed for a {\sc TEMPO} input
parameter file (in the format that {\sc TEMPO} expects), long mode
lists all available parameters and selected mode lists those
parameters which are selected in the check boxes as for normal tabular
output.
 
A system for user feedback is available, with a log being kept of all
comments received.  We greatly appreciate constructive feedback and, provided
the sender's email address is supplied, comments will be acknowledged.

\section{Features for experts}
An
\anchor{http://www.atnf.csiro.au/research/pulsar/psrcat/expert.html}{``expert''
version}\footnote{http://www.atnf.csiro.au/research/pulsar/psrcat/expert.html}
of the web interface provides access to many other parameters in the
catalog database and to many derived parameters which are less
frequently used.  These additional parameters are listed in
Table~\ref{tb:expert}. Additional parameters are displayed on the web
interface with a more compact set of check boxes and additional
documentation is provided for these parameters.

The expert-mode interface also provides for use of custom
databases. The user may upload one or more database files (which must
conform to the {\sc psrcat} data format standard) to the ATNF host
computer. These files may then be used either in place of the public
database file or merged with it. Parameter values in a merged file
overwrite existing values for that parameter and new parameters are
added to the database. Uploaded files may be either deleted at the end
of the session or left for later use. In plots, data from merged files
are highlighted with a heavy cross.

An alternative name may be associated with a pulsar using the {\sf
ALIAS} keyword. Subsequent merged files may use the alternative name
rather than the original name.
 
Four user-defined parameters, {\sf PAR1 - PAR4},
may be included in the uploaded files. They may be accessed, listed,
used in expressions or plotted in the same way as any other parameter.

\subsection{Command-line interface}
All tabular functions of the web interface are available directly on
the command line on linux and unix systems with the program {\sc psrcat}. The
``-h'' option gives a full list of the available options and ``-p''
lists keywords for all available parameters which include and extend the ``expert-mode''
parameters of the web interface. A further argument ``$<$str$>$'' on the
``-p'' option lists only those keywords containing ``$<$str$>$''.  

The current versions of the {\sc psrcat} program and public database
(psrcat.db) may be down-loaded from the ATNF pulsar
\anchor{http://www.atnf.csiro.au/research/pulsar/}{home
page}\footnote{http://www.atnf.csiro.au/research/pulsar/}. The program
is written in the C language and is complete in the sense that no
other libraries are required to compile it. The program makes use of
the freely available
\anchor{http://www.parsifalsoft.com/examples/evalexpression/}{Evaluate}\footnote{http://www.parsifalsoft.com/examples/evalexpression/}
software. \anchor{http://www.gnu.org/}{Gnu}\footnote{http://www.gnu.org/}
<compilers are preferred.

Two environment variables are used by the program: PSRCAT\_FILE
and PSRCAT\_RUNDIR. PSRCAT\_FILE gives the full path to
and name of the default database file; it may be over-ridden using the
``-db\_file $<$path/filename$>$'' option. The ``-all'' option of {\sc
psrcat} merges all files ``obs*.db'' in the PSRCAT\_RUNDIR
directory with the default database file.  Other files may be merged
with the default database file using ``-merge $<$path/filename$>$''
option. Several files can be merged using ``-merge
\verb#"#$<$file1$>$ $<$file2$>$\verb#"#''.  Parameters in
later files overwrite the same parameters in earlier files, including
the main database file.

\subsection{C functions}

Along with the source code for the catalog software, we provide two
simple ``C'' functions that enable a user to obtain catalog parameters
using their own software.  The function ``callPsrcat\_val'' is used
to obtain a numerical parameter value, its error and reference from
the catalog, and ``callPsrcat\_string'' is used to obtain a textual
parameter (such as {\sf SURVEY} or {\sf ASSOC}).  Both functions
require the file name of the catalog (or ``public'' if the
publically available catalog file is to be used), the pulsar name and
the parameter label.  Full descriptions of these routines are
available when downloading the catalog software in a ``README''
file.

\section{Tables and Figures}
The catalog interfaces allow production of many types of parameter
lists. To illustrate this, we give two tables listing relevant
parameters for two categories of pulsars, those with high-energy
(optical, X-ray or gamma-ray) pulsed emission, and those associated
with globular clusters. The web interface also provides facilities for
basic $x - y$ plots and histograms. However, many users will wish to
create files containing custom lists for input into their own plotting
programs or for other manipulation. We give two plots of general
interest based on files produced in this way. Obviously, these figures
and tables represent only a tiny part of what may be produced, but
they illustrate the capabilities of the catalog facility.

Table~\ref{tb:special} lists pulsars of type HE (radio pulsars which
also have detectable high-energy pulsations), type NR (spin-powered
pulsars detectable only at high energies) and type AXP (which includes
pulsating soft gamma-ray repeaters). The table lists database entries
selected by each of the three types, e.g., ``type(nr)'', and displayed
in short format with options ``-nohead -nonumber''. In most cases, the
association was established by the discovery paper; where this is not
the case, the reference key for the paper establishing the association
is given in square brackets. Doubtful associations are followed by
``(?)''. To maintain the requirement that a single entry contains no
spaces, spaces in names of associated objects are replaced by an
underscore.  

\begin{deluxetable}{llrlll}
\tabletypesize{\footnotesize}
\tablecaption{Pulsars of Type AXP, HE or NR\label{tb:special}}
\tablehead{
\colhead{Name} & \colhead{J2000} & \colhead{Period} & \colhead{Age} &
\colhead{$B_s$} & \colhead{Association}\\
\colhead{} & \colhead{Name} & \colhead{(s)} & \colhead{(yr)} &
\colhead{(G)} & }
\startdata
\multicolumn{6}{l}{{\bf Radio pulsars having high-energy pulsations (Type HE):}} \\
J0205+6449 & J0205+6449 & 0.065686 &  5.37e+03 & 3.61e+12 &  SNR:3C58                 \\ 
J0218+4232 & J0218+4232 & 0.002323 &  4.76e+08 & 4.29e+08 &  \nodata                  \\ 
J0437$-$4715 & J0437$-$4715 & 0.005757 &  1.59e+09 & 5.81e+08 &  \nodata                  \\ 
B0531+21   & J0534+2200 & 0.033085 &  1.24e+03 & 3.78e+12 &  SNR:Crab[ccl+69]         \\ 
B0540$-$69   & J0540$-$6919 & 0.050354 &  1.67e+03 & 4.97e+12 &  EXGAL:LMC,SNR:0540$-$693   \\ 
B0656+14   & J0659+1414 & 0.384891 &  1.11e+05 & 4.66e+12 &  SNR:Monogem\_Ring[tbb+03] \\ 
B0823+26   & J0826+2637 & 0.530661 &  4.92e+06 & 9.64e+11 &  \nodata                  \\ 
B0833$-$45   & J0835$-$4510 & 0.089328 &  1.13e+04 & 3.38e+12 &  SNR:Vela                 \\ 
B0950+08   & J0953+0755 & 0.253065 &  1.75e+07 & 2.44e+11 &  \nodata                  \\ 
B1046$-$58   & J1048$-$5832 & 0.123671 &  2.03e+04 & 3.49e+12 &  \nodata                  \\ 
B1055$-$52   & J1057$-$5226 & 0.197108 &  5.35e+05 & 1.09e+12 &  \nodata                  \\ 
J1105$-$6107 & J1105$-$6107 & 0.063193 &  6.33e+04 & 1.01e+12 &  \nodata                  \\
J1124$-$5916 & J1124$-$5916 & 0.135314 &  2.87e+03 & 1.02e+13 &  SNR:G292.0+1.8           \\
B1509$-$58   & J1513$-$5908 & 0.150658 &  1.55e+03 & 1.54e+13 &  SNR:G320.4$-$1.2           \\
J1617$-$5055 & J1617$-$5055 & 0.069357 &  8.13e+03 & 3.10e+12 &  \nodata                  \\
B1706$-$44   & J1709$-$4429 & 0.102459 &  1.75e+04 & 3.12e+12 &  SNR:G343.1$-$2.3(?)[mop93] \\
B1800$-$21   & J1803$-$2137 & 0.133617 &  1.58e+04 & 4.28e+12 &  SNR:G8.7$-$0.1(?)[kw90]    \\
B1821$-$24   & J1824$-$2452 & 0.003054 &  2.99e+07 & 2.25e+09 &  GC:M28                   \\
B1823$-$13   & J1826$-$1334 & 0.101466 &  2.14e+04 & 2.79e+12 &  \nodata                  \\
J1930+1852 & J1930+1852 & 0.136855 &  2.89e+03 & 1.03e+13 &  SNR:G54.1+0.3            \\
B1929+10   & J1932+1059 & 0.226518 &  3.10e+06 & 5.18e+11 &  \nodata                  \\
B1937+21   & J1939+2134 & 0.001558 &  2.35e+08 & 4.09e+08 &  \nodata                  \\
B1951+32   & J1952+3252 & 0.039531 &  1.07e+05 & 4.86e+11 &  SNR:CTB80                \\
J2124$-$3358 & J2124$-$3358 & 0.004931 &  3.80e+09 & 3.22e+08 &  \nodata                  \\
J2229+6114 & J2229+6114 & 0.051624 &  1.05e+04 & 2.03e+12 &  \nodata                  \\[10pt]
\multicolumn{6}{l}{{\bf Non-Radio (Type NR) Pulsars:}} \\				    
J0537$-$6910 &   J0537$-$6910 & 0.016115 & 4.98e+03 &  9.20e+11 & EXGAL:LMC,SNR:N157B \\ 
J0633+1746 &   J0633+1746 & 0.237093 & 3.42e+05 &  1.63e+12 & GRS:Geminga         \\ 
J0635+0533 &   J0635+0533 & 0.033856 & \nodata  &  \nodata  & OPT:Be$-$star         \\
J1210$-$5209 &   J1210$-$5209 & 0.424129 & 3.36e+05 &  2.95e+12 & SNR:G296.5+10.0     \\ 
J1811$-$1925 &   J1811$-$1925 & 0.064667 & 2.33e+04 &  1.71e+12 & SNR:G11.2$-$0.3       \\ 
J1846$-$0258 &   J1846$-$0258 & 0.323598 & 7.22e+02 &  4.85e+13 & SNR:Kes75           \\[10pt]
\multicolumn{6}{l}{{\bf Anomalous X-ray pulsars and Soft gamma-ray repeaters (Type AXP):}} \\
J0100$-$7211 & J0100$-$7211 &  5.439868 &  5.73e+03 & 2.89e+14 & EXGAL:SMC,XRS:CXOU\_J0110043.1$-$721134\\     
J0142+61   & J0142+61   &  8.688330 &  7.02e+04 & 1.32e+14 & XRS:4U\_0142+61                  \\     
J0525$-$6607 & J0525$-$6607 &  8.047000 &  1.96e+03 & 7.32e+14 & SNR:N49(?),SGR\_0526$-$66          \\     
J1048$-$5937 & J1048$-$5937 &  6.452077 &  2.68e+03 & 5.02e+14 & XRS:1E\_1048.1$-$5937              \\     
J1708$-$4008 & J1708$-$4008 & 10.999035 &  8.96e+03 & 4.68e+14 & XRS:1RXS\_J170849.0$-$400910       \\     
J1808$-$2024 & J1808$-$2024 &  7.494782 &  2.81e+02 & 1.80e+15 & SNR:G10.0$-$0.3(?),SGR\_1806$-$20    \\     
J1809$-$1943 & J1809$-$1943 &  5.539220 &  4.26e+03 & 3.42e+14 & XRS:XTE\_J1810$-$197               \\     
J1841$-$0456 & J1841$-$0456 & 11.765730 &  4.51e+03 & 7.06e+14 & SNR:Kes73,XRS:1E\_1841$-$045           \\     
J1845$-$0256 & J1845$-$0256 &  6.971270 &  \nodata  & \nodata  & SNR:G29.6+0.1,XRS:AX\_J1845.0$-$0300   \\     
J1907+0919 & J1907+0919 &  5.168918 &  1.05e+03 & 6.42e+14 & SNR:G42.8+0.6(?),SGR\_1900+14    \\     
J2301+5852 & J2301+5852 &  6.978948 &  2.28e+05 & 5.88e+13 & SNR:CTB109,XRS:1E\_2259.1+586        \\
\enddata
\end{deluxetable}

Globular clusters are rich breeding grounds for millisecond pulsars
due to exchange reactions in the dense cluster core resulting in the
capture of an old neutron star by an evolving star. Subsequent mass
transfer leads to spin-up of the neutron star and reduction in the
effective magnetic field strength and hence a small value of $\dot
P$. Pulsars associated with globular clusters may be extracted from
the catalog using the logical condition ``assoc(gc)'';
Table~\ref{tb:gc_psr} lists some relevant parameters for pulsars
extracted in this way. For many of these pulsars the observed value
of $\dot P$ is negative; this is a consequence of acceleration of the
pulsar in the gravitational field of the cluster
\citep[e.g.,][]{fck+03} and does not represent a speeding up of the
pulsar.  

\begin{deluxetable}{llllrcc}
\tabletypesize{\footnotesize}
\tablecaption{Pulsars in Globular Clusters\label{tb:gc_psr}}
\tablehead{
\colhead{Name} & \colhead{J2000} & \colhead{Association} & \colhead{Period} 
& \colhead{Period} & \colhead{Binary Period} & \colhead{Med. Comp. Mass}\\
\colhead{} & \colhead{Name} & \colhead{} & \colhead{(s)} & \colhead{Derivative} & 
\colhead{(d)} &\colhead{(M$_\sun$)} }
\startdata
B0021$-$72C   &  J0024$-$7204C &  GC:47Tuc  & 0.005757 & $-$4.98e$-$20 &         --  &         --  \\
B0021$-$72D   &  J0024$-$7204D &  GC:47Tuc  & 0.005358 & $-$3.43e$-$21 &         --  &         --  \\
B0021$-$72E   &  J0024$-$7204E &  GC:47Tuc  & 0.003536 &  9.85e$-$20 &    2.2568   &      0.18   \\
B0021$-$72F   &  J0024$-$7204F &  GC:47Tuc  & 0.002624 &  6.45e$-$20 &        --   &        --   \\
B0021$-$72G   &  J0024$-$7204G &  GC:47Tuc  & 0.004040 & $-$4.21e$-$20 &         --  &         --  \\
B0021$-$72H   &  J0024$-$7204H &  GC:47Tuc  & 0.003210 & $-$1.83e$-$21 &     2.3577  &       0.19  \\
B0021$-$72I   &  J0024$-$7204I &  GC:47Tuc  & 0.003485 & $-$4.58e$-$20 &     0.2298  &       0.01  \\
B0021$-$72J   &  J0024$-$7204J &  GC:47Tuc  & 0.002101 & $-$9.79e$-$21 &     0.1207  &       0.02  \\
B0021$-$72L   &  J0024$-$7204L &  GC:47Tuc  & 0.004346 & $-$1.22e$-$19 &         --  &         --  \\
B0021$-$72M   &  J0024$-$7204M &  GC:47Tuc  & 0.003677 & $-$3.84e$-$20 &         --  &         --  \\
B0021$-$72N   &  J0024$-$7204N &  GC:47Tuc  & 0.003054 & $-$2.18e$-$20 &         --  &         --  \\
J0024$-$7204O &  J0024$-$7204O &  GC:47Tuc  & 0.002643 &  3.03e$-$20 &    0.1360   &      0.02   \\
J0024$-$7204P &  J0024$-$7204P &  GC:47Tuc  & 0.003643 &  \nodata  &    0.1472   &      0.02   \\
J0024$-$7204Q &  J0024$-$7204Q &  GC:47Tuc  & 0.004033 &  3.40e$-$20 &    1.1891   &      0.21   \\
J0024$-$7204R &  J0024$-$7204R &  GC:47Tuc  & 0.003480 &  \nodata  &    0.0662   &      0.03   \\
J0024$-$7204S &  J0024$-$7204S &  GC:47Tuc  & 0.002830 & $-$1.20e$-$19 &     1.2017  &       0.10  \\
J0024$-$7204T &  J0024$-$7204T &  GC:47Tuc  & 0.007588 &  2.93e$-$19 &    1.1262   &      0.20   \\
J0024$-$7204U &  J0024$-$7204U &  GC:47Tuc  & 0.004343 &  9.52e$-$20 &    0.4291   &      0.14   \\
J0024$-$7204V &  J0024$-$7204V &  GC:47Tuc  & 0.004810 &  \nodata  &        --   &        --   \\
J0024$-$7204W &  J0024$-$7204W &  GC:47Tuc  & 0.002352 &  \nodata  &    0.1330   &      0.14   \\
J0514$-$4002A &  J0514$-$4002A &  GC:NGC1851& 0.004991 &  \nodata  &   18.7850   &      1.11   \\
B1310+18    &  J1312+1810  &  GC:M53    & 0.033163 &  \nodata  &  255.8000   &      0.35   \\
B1516+02A   &  J1518+0205A &  GC:M5     & 0.005554 &  4.12e$-$20 &        --   &        --   \\
B1516+02B   &  J1518+0204B &  GC:M5     & 0.007947 & $-$3.33e$-$21 &     6.8585  &       0.13  \\
B1620$-$26    &  J1623$-$2631  &  GC:M4     & 0.011076 &  6.70e$-$19 &  191.4428   &      0.33   \\
B1639+36A   &  J1641+3627A &  GC:M13    & 0.010378 &  \nodata  &        --   &        --   \\
J1701$-$3006B &  J1701$-$3006B &  GC:NGC6266& 0.003594 & $-$3.49e$-$19 &     0.1445  &       0.14  \\
J1701$-$3006C &  J1701$-$3006C &  GC:NGC6266& 0.003806 & $-$3.18e$-$20 &     0.2150  &       0.08  \\
J1701$-$3006D &  J1701$-$3006D &  GC:NGC6266& 0.003418 &  \nodata  &    1.1180   &      0.14   \\
J1701$-$3006E &  J1701$-$3006E &  GC:NGC6266& 0.003234 &  \nodata  &    0.1600   &      0.03   \\
J1701$-$3006F &  J1701$-$3006F &  GC:NGC6266& 0.002295 &  \nodata  &    0.2000   &      0.03   \\
B1718$-$19    &  J1721$-$1936  &  GC:NGC6342& 1.004037 &  1.62e$-$15 &    0.2583   &      0.13   \\
J1740$-$5340  &  J1740$-$5340  &  GC:NGC6397& 0.003650 &  1.68e$-$19 &    1.3541   &      0.22   \\
B1744$-$24A   &  J1748$-$2446A &  GC:Ter5   & 0.011563 & $-$3.40e$-$20 &     0.0756  &       0.10  \\
B1745$-$20    &  J1748$-$2021  &  GC:NGC6440& 0.288603 &  4.00e$-$16 &        --   &        --   \\
J1748$-$2446C &  J1748$-$2446C &  GC:Ter5   & 0.008436 & $-$6.06e$-$19 &         --  &         --  \\
B1802$-$07    &  J1804$-$0735  &  GC:NGC6539& 0.023101 &  4.67e$-$19 &    2.6168   &      0.35   \\
J1807$-$2459  &  J1807$-$2459  &  GC:NGC6544& 0.003059 &  \nodata  &    0.0711   &      0.01   \\
B1820$-$30A   &  J1823$-$3021A &  GC:NGC6624& 0.005440 &  3.38e$-$18 &      --     &	    --   \\
B1820$-$30B   &  J1823$-$3021B &  GC:NGC6624& 0.378596 &  3.21e$-$17 &      --     &	    --   \\
B1821$-$24    &  J1824$-$2452  &  GC:M28    & 0.003054 &  1.61e$-$18 &      --     &	    --   \\
B1908+00    &  J1910+0004  &  GC:NGC6760& 0.003619 &  \nodata  &    0.141    &        0.02 \\ 
J1910$-$5959A &  J1910$-$5959A &  GC:NGC6752& 0.003266 &  3.07e$-$21 &    0.837    &        0.22 \\ 
J1910$-$5959B &  J1910$-$5959B &  GC:NGC6752& 0.008358 & $-$7.99e$-$19 &      --     &	    --   \\
J1910$-$5959C &  J1910$-$5959C &  GC:NGC6752& 0.005277 &  2.20e$-$21 &      --     &	    --   \\
J1910$-$5959D &  J1910$-$5959D &  GC:NGC6752& 0.009035 &  9.63e$-$19 &      --     &	    --   \\
J1910$-$5959E &  J1910$-$5959E &  GC:NGC6752& 0.004572 & $-$4.37e$-$19 &      --     &	    --   \\
B2127+11A   &  J2129+1210A &  GC:M15    & 0.110665 & $-$2.10e$-$17 &      --     &	    --   \\
B2127+11B   &  J2129+1210B &  GC:M15    & 0.056133 &  9.56e$-$18 &      --     &	    --   \\
B2127+11D   &  J2129+1210D &  GC:M15    & 0.004803 & $-$1.07e$-$17 &      --     &	    --   \\
B2127+11E   &  J2129+1210E &  GC:M15    & 0.004651 &  1.78e$-$19 &      --     &	    --   \\
B2127+11F   &  J2129+1210F &  GC:M15    & 0.004027 &  3.20e$-$20 &      --     &	    --   \\
B2127+11G   &  J2129+1210G &  GC:M15    & 0.037660 &  2.00e$-$18 &      --     &	    --   \\
B2127+11H   &  J2129+1210H &  GC:M15    & 0.006743 &  2.40e$-$20 &      --     &	    --   \\
B2127+11C   &  J2130+1210C &  GC:M15    & 0.030529 &  4.99e$-$18 &    0.335    &       1.13  \\
J2140$-$2310A &  J2140$-$2310A &  GC:M30    & 0.011019 & $-$5.18e$-$20 &    0.170    &       0.11  \\
J2140$-$23B   &  J2140$-$23B   &  GC:M30    & 0.012986 &  \nodata  &      --     &	     --  \\
\enddata    
\end{deluxetable}

\begin{figure}
\centerline{\psfig{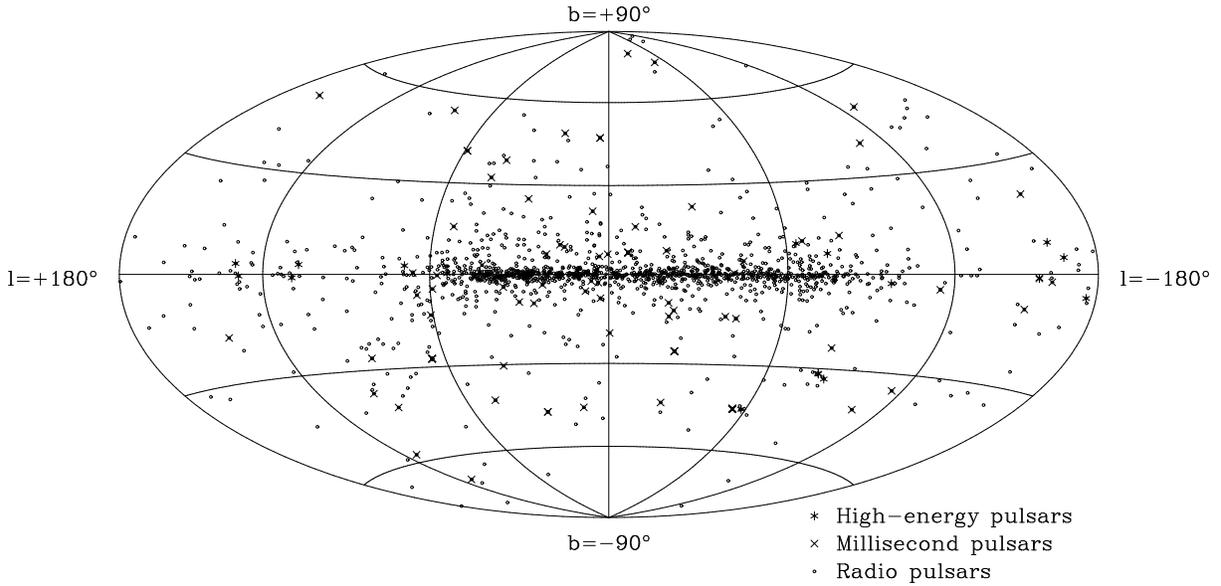}}
\caption{Distribution of pulsars on an Hammer-Aitoff equal-area
projection in Galactic coordinates with the Galactic Center at the
center of the plot.}\label{fg:gal_lb}
\end{figure}

With either the web or command-line interfaces, it is simple to
produce lists of parameters and to copy these to a file to be used as
input to other programs for custom plotting or other purposes. As an
example, Fig.~\ref{fg:gal_lb} shows the distribution of all known
pulsars in Galactic coordinates. Most high-energy pulsars are young
(median characteristic age $\sim 2\times 10^4$~yr) and hence are
concentrated along the Galactic plane, whereas most millisecond
pulsars are very old (median characteristic age $\sim 4\times
10^9$~yr) and have therefore had time to migrate away from their
region of birth. They are therefore more widely distributed in
Galactic latitude.

\begin{figure}
\centerline{\psfig{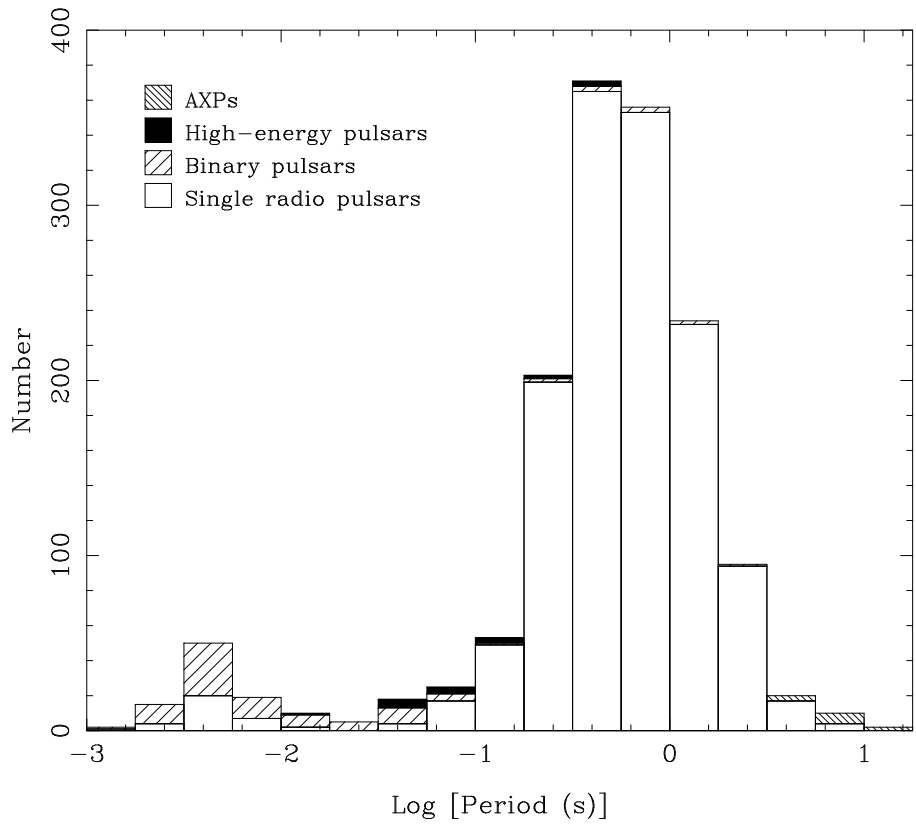}}
\caption{Distribution of pulse periods for all known pulsars, with
  binary pulsars, spin-powered pulsars with high-energy (optical,
  X-ray or gamma-ray) pulsed emission and AXPs separately
  identified.}\label{fg:phist}
\end{figure}

As another example, we show in Fig.~\ref{fg:phist} a histogram of the
distribution of pulsar periods for all known pulsars, divided into
binary pulsars, high-energy pulsars, AXPs and single radio pulsars
using the {\sf Type} keyword. This plot shows the clear dichotomy
between millisecond pulsars and so-called ``normal'' pulsars. Binary
pulsars predominantly have periods in the millisecond range whereas
all AXPs are at the other end of the histogram with periods in the
range 5 -- 12~s. High-energy emitters are generally young and most
have periods in the range 30 -- 150~ms.

\section{Conclusions}
We have compiled an up-to-date pulsar catalog based on data from
published papers and developed web and command-line interfaces to
access both the catalog data and parameters derived from them. Full
bibliographic information is provided for all data contained in the
catalog.  Supporting documentation and a mechanism for user feedback
are also provided. Both the database and the software associated with
the command-line interface are freely available for research
purposes. The catalog will be updated at intervals to include recently
published material and to correct any errors brought to our
attention. An ``expert-mode'' web interface is also provided, which
gives access to a wider range of parameters and allows use of custom
databases. 

\section*{Acknowledgments}
Many people have contributed to the maintenance and upgrading of the
database used for the Taylor, Manchester \& Lyne (1993) paper. We
particularly thank Andrew Lyne of the University of Manchester,
Jodrell Bank Observatory, David Nice of Princeton University and
Russell Edwards, then at Swinburne University of Technology. We also
acknowledge the efforts of Warwick University students Adam Goode and
Steven Thomas who compiled and checked a recent version of the
database. The glitch database is jointly maintained by Jodrell Bank
Observatory and the ATNF. This work has made extensive use of NASA's
Astrophysics Data System, the Google search engine and the PGPLOT
plotting package. Finally, we thank our colleagues for their comments
and suggestions which have helped to improve both the database and the
means of accessing it.


\end{document}